\documentclass[article,twocolumn,prd,superscriptaddress,preprintnumbers,nofootinbib]{revtex4-2}
\usepackage[utf8]{inputenc}
\usepackage{amsmath,amsthm,amssymb,mathtools,physics,cancel,mhchem,tikz}
\usepackage{graphicx,flexisym,float,mwe}
\usepackage{dcolumn}
\usepackage[hidelinks]{hyperref}
\hypersetup{colorlinks=true,
    linkcolor=blue,
    filecolor=blue,
    urlcolor=blue,
    citecolor=blue,
    }
\usepackage{caption}
\usepackage{subcaption}
\usepackage{adjustbox}
\usepackage{gensymb}
\usepackage{breqn}
\usepackage{braket}
\usepackage{enumitem}
\usepackage{gensymb}
\usepackage{mhchem}
\graphicspath{{./figures/}}

\begin{document}
\begin{flushright}
MI-HET-XXX
\end{flushright}
\title{Novel approach to investigate ATOMKI anomaly using Coherent CAPTAIN-Mills detectors}

\begin{abstract}
ATOMKI nuclear anomaly has suggested a new BSM (Beyond the Standard Model) boson with mass $\sim17$ MeV emitted from excited nuclei and quickly decays into a pair of $e^+e^-$. In order to search for the new particle, we propose a new approach that utilizes the ongoing Coherent CAPTAIN-Mills (CCM) 10-ton LAr (liquid argon) detectors.
The neutrons from the Lujan target can scatter inelastically by the PMT glass in the CCM detector can produce the new boson which solves the ATOMKI anomaly.
The new boson can be detected from its decay to a $e^+e^-$ pair.
We find that CCM probes a large area of the anomaly-allowable parameter space. We also show the prediction for a 100-ton LAr detector and 5-ton EOS water detector.
\end{abstract}

\author{Bhaskar Dutta}
\email{dutta@physics.tamu.edu}
\affiliation{Mitchell Institute for Fundamental Physics and Astronomy$,$ Department~ of ~Physics ~ and~ Astronomy$,$\\ Texas A$\&$M University$,$ ~College~ Station$,$ ~Texas ~77843$,$~ USA}

\author{Bai-Shan Hu}
\email{baishan@tamu.edu}
\affiliation{Cyclotron Institute and Department of Physics and Astronomy$,$\\ Texas A$\&$M University$,$ ~College~ Station$,$ ~Texas ~77843$,$~ USA}
\affiliation{National Center for Computational Sciences and Physics Division$,$ ~Oak ~Ridge ~National ~Laboratory$,$ ~Oak ~Ridge$,$ ~Tennessee ~37831$,$ ~USA}

\author{Wei-Chih Huang}
\email{s104021230@tamu.edu}
\affiliation{Mitchell Institute for Fundamental Physics and Astronomy$,$ Department~ of ~Physics ~ and~ Astronomy$,$\\ Texas A$\&$M University$,$ ~College~ Station$,$ ~Texas ~77843$,$~ USA}

\author{Richard G.~Van~de~Water}
\email{vdwater@lanl.gov}
\affiliation{Los ~Alamos ~National ~Laboratory$,$ ~Los ~Alamos$,$ ~NM ~87545$,$ ~USA}

\maketitle

{\bf{Introduction}}
Beyond the Standard Model new physics ideas are well-motivated. Various ongoing experiments are probing scales of new physics ideas that could explain the origin of dark matter, neutrino masses, mixings, and anomalies observed in different experiments. Rare decay processes, various electron and proton beam dump experiments, reactor-based experiments, and astrophysical observations are probing models possessing low energy scales for new physics while the LHC is searching for higher energy scales mostly around and above the weak scale.  A wide range of new physics scales are also being searched at the direct and indirect detection experiments.

Among the existing anomalies, a very interesting excess has been reported by the ATOMKI pair spectrometer experiment~\cite{Feng:2016jff,Cartlidge2016,Krasznahorkay:2015iga,Krasznahorkay_2018} in recent times. A more than 5 $\sigma$ excess is observed in the internal pair conversion (IPC) decays of excited $^8$Be nuclei with noticeable bumps in both the invariant mass and the angular spectra of $e^+e^-$. These bumps cannot be explained by electromagnetic IPC from virtual photons where a smooth and rapidly declining distribution of opening angles is expected. More recently, similar anomalous measurements are reported in the IPC decays of excited $^4$He~\cite{PhysRevC.104.044003,Krasznahorkay:2019lyl} and $^{12}$C nuclei ~\cite{Krasznahorkay:2022pxs}.

The anomalies suggest there could be a new boson $X$ with mass $\sim17$ MeV \cite{Barducci:2022lqd,PhysRevC.104.044003,Krasznahorkay:2019lyl,Krasznahorkay_2018,Krasznahorkay:2015iga,Krasznahorkay:2022pxs,Feng:2016ysn,Feng:2020mbt,Kozaczuk:2016nma} in the de-excitations of the nuclei such as $^8$Be, $^4$He, and $^{12}$C. The decay of the slow-moving boson $X$ into a pair of $ e^+e^-$ explains the anomaly where
the boson $X$ requires to have interactions with quarks and $e^+e^-$. Since $X$ decays quickly into the $e^+e^-$ pair, the coupling cannot be too small and it is found that coupling strengths $\sim 10^{-3} - 10^{-4}$ can explain the excess.

At present, several experimental groups are trying to probe this excess by performing similar measurements as ATOMKI. 
In this paper, we propose a novel way to investigate this anomaly at the ongoing proton beam-dump based neutrino experiments. We utilize the ongoing 1 GeV proton beam-based experiments, e.g., CCM~\cite{PhysRevD.106.012001,CCM:2021yzc,CCM:2021jmk,Aguilar-Arevalo:2023kvr}, where the detectors are situated around 20 meter from the Lujan target.
We will also include a 100-ton upgrade (PIP2-BD) in our analysis for a future projection~\cite{Toups:2022yxs}, and the 5-ton EOS detector if it is situated at Lujan~\cite{Anderson:2022lbb}.

The $X$ boson, in our proposal, needs to be produced and decay in the detector, since its lifetime is small compared to the time of flight from the target to the detector. We utilized excited states of oxygen around 18 MeV and above with appropriate isospins to produce the $X$ boson.

In CCM, the photomultiplier tubes (PMT) glass is made out of SiO$_2$ and there are 200 of them in the inner walls of the detector.
The PMT glass oxygen transitions to the excited state by inelastic scattering of neutrons from the target.
The LAr used in the CCM detector does not possess any excited state around 17 MeV and above with sufficient strength to produce the $X$ boson.
We will calculate the neutron-oxygen inelastic cross-sections using the nuclear Gamow shell model to estimate the production of a 17 MeV $X$ boson. 
The final state electrons emerging from the $X$ decay are of $\mathcal{O}$(10) MeV. 

Nevertheless, there exists literature \cite{MEGII:2024urz,Zhang:2020ukq} that contradicts ATOMKI anomaly.
In which \cite{MEGII:2024urz} finds no significant signal of $X$ boson. 
Ref.~\cite{Zhang:2020ukq} claims that the $X$ boson production is dominated by direct transitions such as $E1$, instead of nuclear de-excitation and that protophobic vector boson $X$ does not explain ATOMKI anomaly. 

\begin{table*}[th]
    \centering
    \caption{Summary of variables in Eq.~\ref{eq:gamma_flux}.}
    \begin{tabular}{c|c|c|c}
    \hline
    Variable & Definition & CCM & PIP2-BD \\ \hline
    $\phi_n$ & Neutron flux & \multicolumn{1}{c}{$\mathcal{O}(10^{-4}) - \mathcal{O}(1)$ cm$^{-2}$s$^{-1}$}\\ \hline
    $\sigma_n$ & Neutron cross section & 
    \multicolumn{2}{c}{$\mathcal{O}(10^{-3}) - \mathcal{O}(10^{-2})$ barn} \\ \hline
    $M_s$ & PMT glass molar mass & \multicolumn{1}{c}{60.08 ($\ce{SiO_2}$)} \\ \hline
    $n$ & Number of nucleus per molecule & \multicolumn{1}{c}{2 ($\ce{SiO_2}$)}  \\ \hline
    $w$ & PMT glass window weight & 130 kg & 841.1 kg  \\ \hline
    $\Gamma_{\text{tot}}$ & $\mathcal{N}^* \rightarrow \mathcal{N} +\text{any}$: Total decay width & \multicolumn{1}{c}{$\mathcal{O}(10)-\mathcal{O}(10^2)$ keV} \\ \hline
    $\Gamma_\gamma$ & $\mathcal{N}^* \rightarrow \mathcal{N} +\gamma$: Photon decay width & \multicolumn{1}{c}{$\mathcal{O}(10^{-2}) - \mathcal{O}(10^{-1})$ keV} \\ \hline
    $N_A$ & Avogadro constant & \multicolumn{1}{c}{$6\times 10^{23}$} \\ \hline
    \end{tabular}\label{tab:symbols}
\end{table*}

{\bf{Model}} Over the last several years, various models have been proposed to explain the anomaly and it appears that new interactions involving an axial vector boson appear to explain the anomaly after satisfying all the experimental constraints.

The Lagrangian of such interaction is given by
\begin{equation}
    \mathcal{L} = \epsilon_p \bar{p} \gamma^\mu \gamma^5 p X_\mu + \epsilon_n \bar{n} \gamma^\mu \gamma^5 n X_\mu
\end{equation}

\begin{figure}[tbh]
    \centering
    \includegraphics[width=\columnwidth]{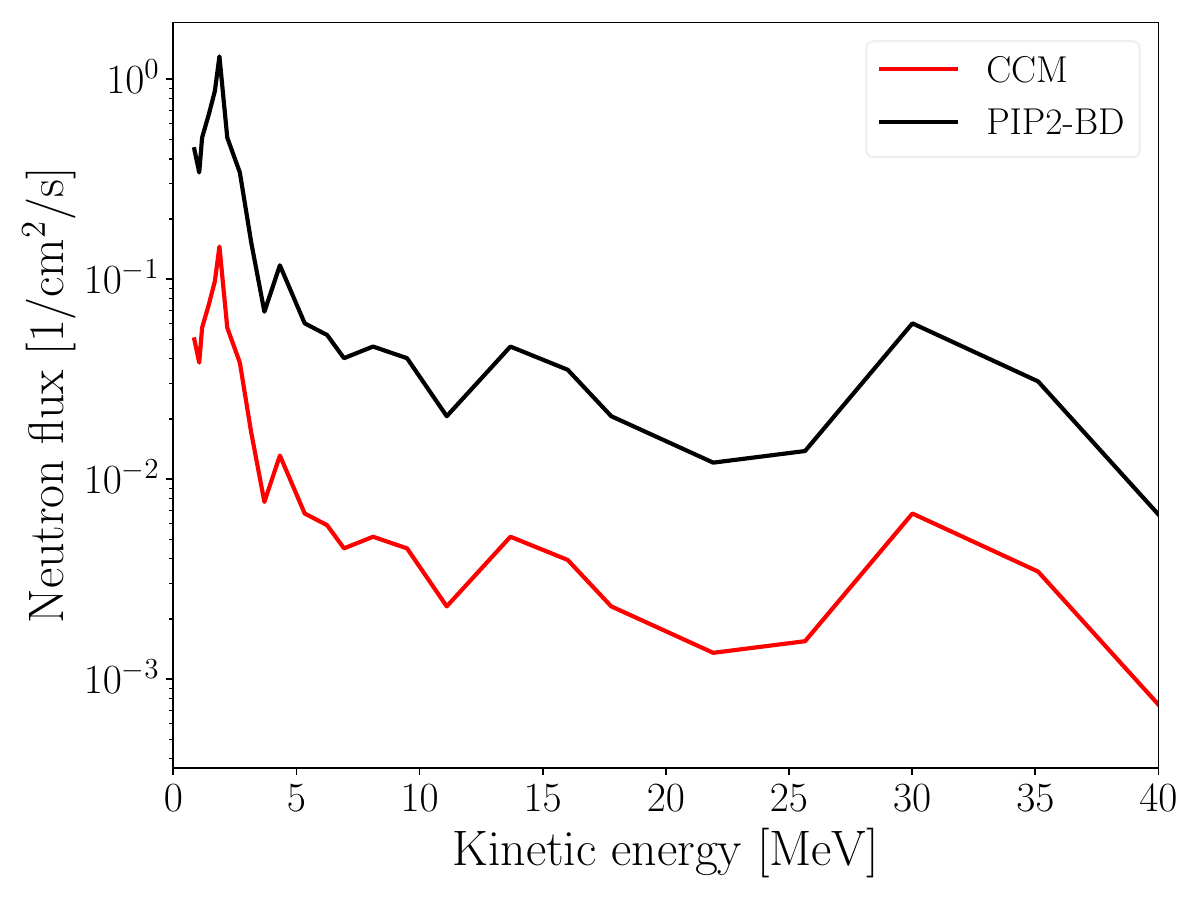}
    \caption{Neutron flux at CCM LAr detector (after the shielding) and PIP2-BD LAr 100t LAr detector.}
    \label{fig:n_flux}
\end{figure}

{\bf{Production of $X$}} We utilize the neutron flux produced at the target when the $\sim$ 1 GeV proton beam hits the target.
The neutron that arrives at the CCM detector and PIP2-BD detector is shown in Fig.~\ref{fig:n_flux}.
The CCM flux is computed using the MCNP simulation incorporating the target geometry which matches with the experimental data~\cite{Richard:2023int}. The neutron flux at PIP2-BD is assumed to be scalable from CCM by distance and POT (protons-on-target).

In CCM and PIP2-BD experiments, the neutron excites the oxygen in the PMT glass window~\cite{r5912}.
The PMT glass window is made of borosilicate glass ($\ce{SiO_2}$), in which $^{28}$Si does not have any excitation levels above 17 MeV in current experiments, while $^{16}$O has several levels above 17 MeV~\cite{TILLEY19931}.
$X$ can be produced during the deexcitation of the nucleus.
The excited nuclei decay to the
ground state and emit a photon or $X$.
Shortly after $X$ is produced, it decays to $e^\pm$, which will be detected by the PMT dynode.
The production and decay of $X$ is as follows:
\begin{align}
n + \mathcal{N} & \rightarrow n + \mathcal{N}^* \\
\mathcal{N}^* & \rightarrow \mathcal{N} + \gamma / X \label{eq:decay} \\
X & \rightarrow e^+ + e^-
\end{align}
where $\mathcal{N}=\,^{16}$O.

The minimum of opening angle of $e^\pm$ is given by
\begin{equation}
    \theta^{\rm min}_{\pm} \approx \cos^{-1}(2v_X^2-1)
\end{equation}
where $v_X$ is velocity of $X$ boson. The opening angle is generally close to $\pi$, which induces a unique signature of the $X$ boson. This unique signature enhances background subtraction and thus improves the sensitivity of the search.

{\bf{Benchmark experiments}}

CCM makes use of a 0.8 GeV proton beam (0.29$\mu$s wide, 20 Hz frequency) impinging on a tungsten target, which produces $5.6\times10^{14}$ (protons-on-target) POT per second.
Currently, CCM is operating a 10-ton (7 ton fiducial) LAr detector, located 20 meter away from the target.
The proton beam impinging on the tungsten target produces a large amount of neutrons (about 20 neutrons per proton).
To reduce the neutron and random backgrounds for the primary physics goals, CCM has concrete and steel shielding surrounding the LAr detector. This reduces the neutrons from the target, which will negatively impact the ATMOKI analysis. The neutron flux after shielding is shown in Fig.~\ref{fig:n_flux}.

PIP2-BD at Fermilab~\cite{Toups:2022yxs,Pellico:2022dju,Aguilar-Arevalo:2023dai} is a future experiment and will use a 2 GeV proton beam (2$\mu$s wide, 120 Hz frequency) impinging on a light target such as carbon.
A 100-ton LAr detector with a 20 keV threshold will be located at different distances, 15 m or 30 m, from the target. Similar experiments are also being proposed at LANL.

\begin{table*}[th]
    \centering
    \begin{tabular}{c|c|c|c|c|c}
    \hline
    Nucleus & Energy [MeV $\pm$ keV] & $J^\pi$ & $\Gamma_{\text{tot}}$ [keV] & $\hat{O}$ & $|\langle J_i |\hat{O}| J_f \rangle |^2$ \\ \hline
    $^{16}$O & 17.09 $\pm$ 40 & $1^-$ & 380 $\pm$ 40 & $\hat{D^\sigma_3}$ & $7.29310\times 10^{-6}$ \\ \hline
    $^{16}$O & 17.14 $\pm$ 10& $1^+$ & 34 $\pm$ 3 & $\hat{\sigma}$ & 0.042008 \\ \hline
    $^{16}$O & 17.282 $\pm$ 11& $1^-$ & 78 $\pm$ 5 & $\hat{D^\sigma_3}$ & $7.29310\times 10^{-6}$ \\ \hline
    $^{16}$O & 18.79 $\pm$ 10& $1^+$ & 120 $\pm$ 20 & $\hat{\sigma}$ & 0.011664 \\ \hline
    $^{16}$O & 19.47 $\pm$ 30& $1^-$ & 200 $\pm$ 70 & $\hat{D^\sigma_3}$ & $2.38115\times 10^{-5}$ \\ \hline
    $^{16}$O & 20.945 $\pm$ 20& $1^-$ & 300 $\pm$ 10 & $\hat{D^\sigma_3}$ & $3.02156\times 10^{-6}$ \\ \hline
    $^{16}$O & 22.15 $\pm$ 10& $1^-$ & 680 $\pm$ 10 & $\hat{D^\sigma_3}$ & $1.87670\times 10^{-6}$ \\ \hline
    $^{16}$O & 22.89 $\pm$ 10 & $1^-$ & 300 $\pm$ 10 & $\hat{D^\sigma_3}$ & $1.87670\times 10^{-6}$ \\ \hline
    \end{tabular}
    \caption{Available nuclear states, where $\hat{O}$ is the non-vanishing operator in the $X$ decay rate. 
    $|\langle J_i |\hat{O}| J_f \rangle |^2$ is computed by shell model, the other columns are experimental measurements. $|\langle J_i |\hat{\sigma}| J_f \rangle |^2$ is dimensionless, $|\langle J_i |\hat{D^\sigma_3}| J_f \rangle |^2$ is in MeV$^{-2}$.}
    \label{tab:states}
\end{table*}

{\bf{Analysis}}
$X$ boson flux $\phi_X$ can be calculated from the transition lines of the nucleus $\mathcal{N}=\,^{16}$O as follows:
\begin{equation}
    \label{eq:x_flux}
    \phi_X = \phi_\gamma \frac{\Gamma_X}{\Gamma_\gamma}
\end{equation}
where $\Gamma_\gamma$ and $\Gamma_X$ are decay rates for $\mathcal{N}^* \rightarrow \mathcal{N}+\gamma$ and $\mathcal{N}^* \rightarrow \mathcal{N}+X$, respectively. Decay rates $\Gamma_X$ for $1^+$ states and $1^-$ states are given by \cite{Barducci:2022lqd}
\begin{align}
\label{eq:decay_rate}
\Gamma^{J=1^+}_X &= {p_X \over 18\pi} \left(2+ {E_X^2 \over m_X^2}\right) \lvert \langle \epsilon_p \hat{\sigma}^p + \epsilon_n \hat{\sigma}^n \rangle \rvert^2 \nonumber \\
&= {p_X \over 18\pi} \left(2+ {E_X^2 \over m_X^2}\right) (\epsilon_p + \epsilon_n)^2 \lvert \langle \hat{\sigma} \rangle \rvert^2 \\
\Gamma^{J=1^-}_X &= {p_X^3 \over 144\pi} \left( \epsilon_p - \epsilon_n \right)^2 \lvert \langle \hat{D^\sigma_3} \rangle \rvert^2 
\end{align}
where $\hat{\sigma}$ is spin operator, $\hat{D^\sigma_3}$ is isovector axial spin dipole. We assume isosymmetry, i.e. $\hat{\sigma}^p=\hat{\sigma}^n$. Refer to the Appendix for derivation.

The photon flux, $\phi_\gamma$, is calculated  from the inelastic neutron scattering is given as:
\begin{align}
    \phi_\gamma = n w {N_A \over M_s} {\Gamma_\gamma \over \Gamma_{\text{tot}}} \int \sigma_n(E_n) {d\phi_n \over dE_n} dE_n
    \label{eq:gamma_flux}
\end{align}
where $\Gamma_\gamma$ is with a certain excited state of $\mathcal{N}$. The detailed definitions and values of the other variables are described in Table.~\ref{tab:symbols}. Note that $\Gamma_\gamma$ cancel out by combining Eq.~\ref{eq:x_flux} and Eq.~\ref{eq:gamma_flux} and that the $X$ boson flux is independent of the photon decay branching ratio ${\Gamma_\gamma \over \Gamma_{\text{tot}}}$ but dependent on the total decay width $\Gamma_{\text{tot}}$.

{\bf{Nuclear physics}}
Given that $X$ boson is spin-1 and $^{16}$O ground state is spin-0, the nuclear excited states must be spin-1. The energy of excited states must also be at least 17 MeV due to the mass of $X$.
Both positive and negative parities are feasible.
We consider up to $\sim 23$ MeV of state energy since the higher energy states would likely have lower neutron cross section.
All the available states \cite{KELLEY201771,TILLEY19931} are summarized in Table.~\ref{tab:states}.

We perform a large-scale shell model (LSSM) to compute the strength $|\langle J_i |\hat{O}| J_f \rangle |^2$ using the code BIGSTICK \cite{Johnson:2018hrx,Johnson:2013bna}. The YSOX interaction ~\cite{PhysRevC.46.923,Yuan:2012zz} is exactly diagonalized within full $psd$ model space. Spin operator $\hat{\sigma}$ ($1^+$ states) is relatively easy to calculate.
However the computation of $\hat{D^\sigma_3}$ ($1^-$ states) is challenging for $^{16}$O.
The energy gaps between the states are small in both of shell model calculation and experimental measurement, so it's difficult to uniquely identify the mapping of a calculated state to a measured state.
Hence, there are some $1^-$ $^{16}$O states share the same calculated $\hat{D^\sigma_3}$ strength.

{\bf{Neutron cross section}}
We calculate the cross section of inelastic neutron scattering reactions $^{16}$O$(n,n^{\prime})$ using the state-of-the-art coupled-channels Gamow shell model (GSM-CC) \cite{michel2021gamow,PhysRevC.106.L011301}.
Employing the core + valence configuration-interaction framework, the GSM-CC approach is the tool par excellence for studies of both nuclear structure and reaction observables at the same time.
In this work, the $^{12}$C core for target nucleus $^{16}$O is mimicked by a Woods-Saxon potential, and the interaction among valence nucleons is modeled through the Furutani-Horiuchi-Tamagaki (FHT) nuclear interaction \cite{PTP.62.981,PTPS.68.193,PhysRevC.106.L011301}, in which the Coulomb interaction is added for protons. 

To investigate uncertainties due to nuclear Hamiltonians, we perform three different GSM-CC models for $^{16}$O$(n,n^{\prime})$ calculations.
For the $^{16}$O$(n,n^{\prime})$ reaction, the first was established from Ref.~\cite{PhysRevC.106.L011301}, which has been successfully used for the description of the reaction $^{15}$O$(p,p)$ and $^{14}$O$(p,p)$. In this calculation, neutron $psd$ partial waves are represented using the Berggren basis, which treats bound states, unbound resonant states, and nonresonant continuum states on an equal footing. While the harmonic oscillator (HO) states are used for the proton $psd$ partial waves.
Details of the calculation are similar to those elaborated in Ref.~\cite{PhysRevC.106.L011301}. The second is based on the $psd$-shell GSM interaction \cite{PhysRevC.96.054316}.
It gives comprehensive studies of structure and reactions aspects of nuclei from the $psd$ region of the nuclear landscape.
The third one fits low-lying spectra of $^{13-15}$C, $^{13-16}$N, $^{14,15}$O and $^{15,16}$F, and includes the $sdpf$ partial waves \cite{Jianguo}.


\begin{figure}[tbh]
    \centering
    \includegraphics[width=\columnwidth]{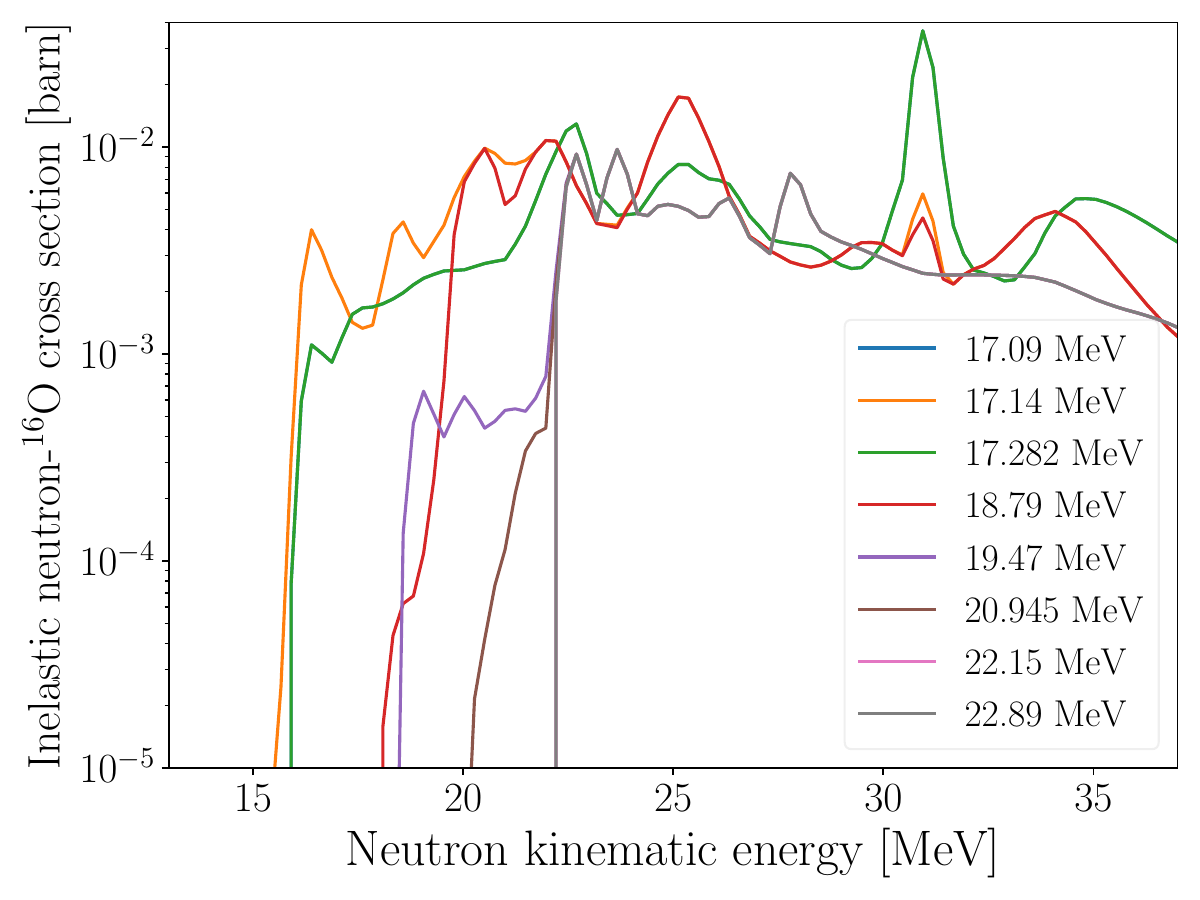}
    \caption{Inelastic neutron-nucleus cross section for $^{16}$O. The labels indicate the energy of the state.}
    \label{fig:n_xsec}
\end{figure}

The results of neutron cross section among different shell models we calculate are consistent and within one order of magnitude of each other.
However, these different GSM-CC calculations still have discrepancies with experimental measurements for the high excitation energies relevant to this work. 
To incorporate this uncertainty into the neutron cross section, we set an energy tolerance ({\it e.g.}, $\pm$ 3 MeV) for each experiment state.
The calculated states which have the matching parity and stay within the tolerance are selected.
Finally, the neutron cross section of the experiment state is the maximum of the calculated cross sections of the selected states, as shown in Fig.~\ref{fig:n_xsec}

{\bf{Parameter space search}}
Summing over the states in Table~\ref{tab:states}, the expected number of $X$ boson events
\begin{equation}
    N_X= \Delta T \sum_i \phi_X^i
\end{equation}
where $\Delta T$ is the observation time, $\phi_X^i$ is $X$ flux produced by nuclear state $i$.

\begin{figure}[tbh]
    \centering
    \includegraphics[width=\columnwidth]{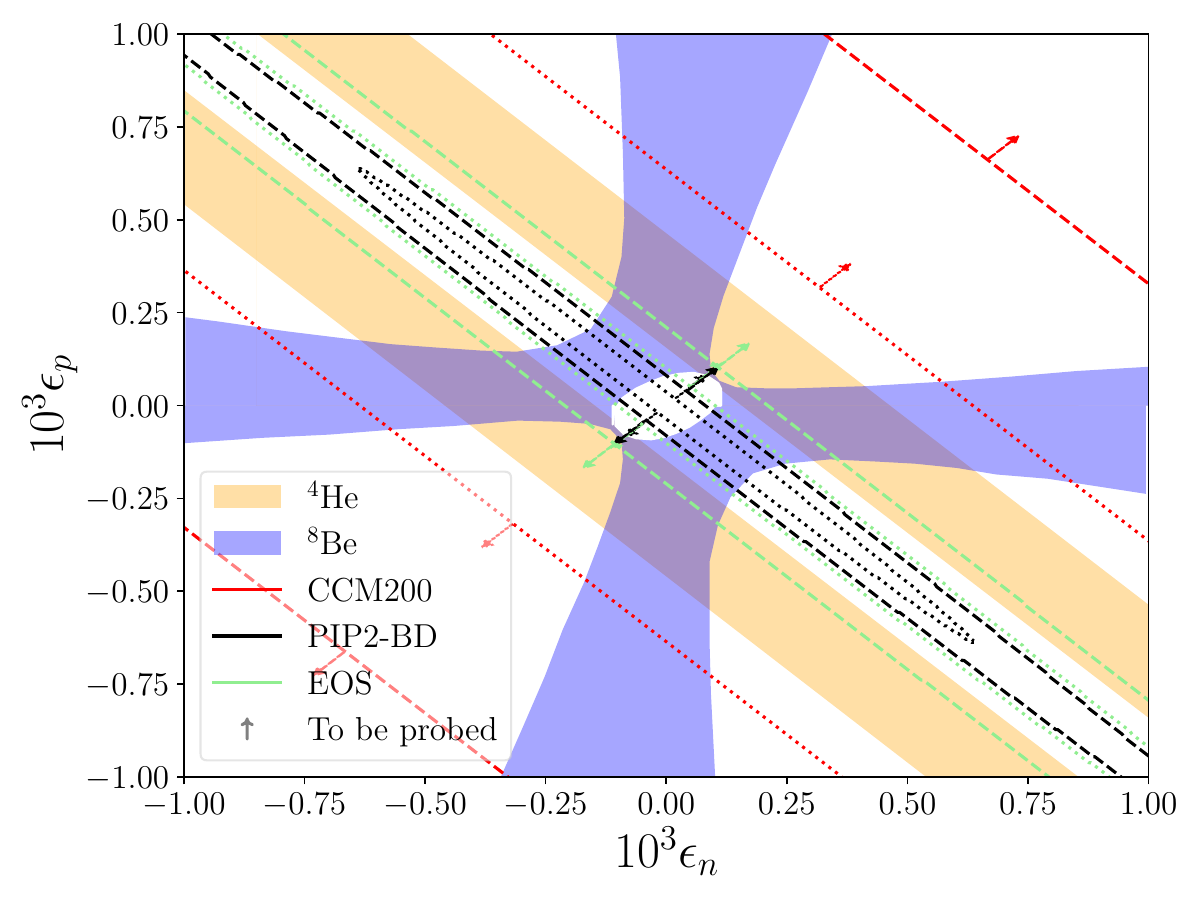}
    \caption{The disallowed space (indicated by arrows) of ATOMKI for 2.3 events (dotted line) and 10 events (dashed line) for CCM (red), PIP2-BD (black) and EOS (green). The color shaded region is the allowed space of $^{4}$He (yellow) and $^{8}$Be (blue). $\epsilon_p$ and $\epsilon_n$ are coupling of $X$ boson to proton and neutron, respectively.}
    \label{fig:limits}
\end{figure}

We investigate the $X$ boson in CCM and PIP2-BD.
CCM has 200 PMTs (total of 69.33\,kg of oxygen in the 130\,kg of glass), $7.5\times 10^{21}$ POT per year, and has begun a three year run ending in 2025.
PIP2-BD has 1294 PMTs, $9.9\times 10^{22}$ POT per year, and plans to run for at least 5 years.
The PMT $e^+e^-$ signal deposits most of its energy as scintillation light in the glass, producing events with very unique reconstruction signatures. CCM data analysis has shown that the scintillation backgrounds above 10\,MeV similar to $e^+e^-$ signal process is small~\cite{CCM:2021jmk}. However, it is prudent to further reduce the backgrounds from the many neutron inelastic scattering channels producing protons, deuterons, alpha particles on oxygen, and other in-situ materials and contaminants.   To achieve this   reconstruction of the opening angle and energy of the $e^+e^-$ pair from the $X$ decay would enable determination of the invariant mass, further restricting the backgrounds.  Reconstruction of the invariant mass in the PMT glass will be difficult since the geometry of the $e^+e^-$ interaction is not ideal, with a high probability of one of the leptons escaping the glass without depositing much energy.  To improve invariant mass reconstruction, we plan to deploy a low contaminate solid spherical piece of pure glass with a weight on the order of 130\,kg (similar to all the PMT glass) at the center of the CCM detector.  This will provide a much more favorable $e^+e^-$  geometry to reconstruct the lepton pair opening angle and individual energies (see Appendix C).  Recent work on CCM has shown the ability to reconstruct Cherenkov light down to a $\sim$MeV for $\gamma$-rays from a Na-22 source~\cite{Darcy:2024pheno}, which will further improve invariant mass reconstruction.  The reconstruction of Cherenkov light is critical to reconstruct the individual $e^+e^-$ energies and angle in order to reconstruct the invariant mass.  A benefit of using Cherenkov light reconstruction is that it rejects final state backgrounds from neutron inelastic scattering which only produces scintillation light. 
Fig.~\ref{fig:limits} shows the limits of our model on the ATOMKI anomaly at CCM and PIP2-BD with 2.3 and 10 signal events. The shaded region (blue and orange) emerges from the ATOMKI signal associated with $^8$Be and $^4$He, respectively~\cite{Barducci:2022lqd}.
It's worth mentioning that it's challenging to simultaneously fit all examined nuclei ($^{16}$O, $^8$Be, and $^4$He) due to the uncertainties of nuclear physics such as energy, transition strength, and operator matrix element.
We find that CCM investigate the parameter space associated with the signal in a complementary way and a large portion of the parameter space will be probed.
The proposed PIP2-BD detector would cover the remaining available parameter space. Finally, sensitivities are also shown for the 5-ton water (H$_{2}$O) based EOS detector~\cite{Anderson:2022lbb} where assuming it is situated at Lujan at the same distance and collects that same POT as CCM.  EOS will have excellent Cerenkov light reconstruction and better $X$ particle invariant mass reconstruction than CCM.  With the increase in target mass, the sensitivities for EOS will be orders of magnitude better and will cover almost the entire parameter space.

{\bf{Conclusions}}
We investigated the ATOMKI anomaly in a novel way in stopped pion facilities, e.g., ongoing CCM,  planned PIP2-BD and EOS.
We take advantage of the neutrons produced when the proton beam with $\sim 1$ GeV hits the target. Neutrons are usually treated as background. However, to address the ATOMKI anomaly, we utilize these neutrons, which arrive at the detector.
The neutrons inelastically scatter, excite the nucleus in the PMT glass, and produce the $X$ boson which quickly decays into a pair of $e^+e^-$. We calculate the neutron-oxygen inelastic cross-sections using the neutron energy spectra at CCM.
The nuclear shell model produces a reliable estimation of neutron cross-section. 
Using the calculations of the shell model along with the experimental measurements of the strengths, we can estimate the flux of the $X$ boson from nuclear deexcitation processes.
We perform the searches in ongoing and future experiments.  
The ongoing experiment, CCM, excludes a large area of signal parameter space, while future PIP2-BD and EOS experiments can exclude the remainder of the available parameter space.
Beyond the investigation of the ATOMKI anomaly, the production of a $X$ boson with mass $\lesssim 20$ MeV from the inelastic scattering of neutrons can be used to investigate various low-scale models.

\begin{acknowledgments}
We thank Nicolas Michel for providing us with the GSM-CC code for calculating neutron cross section, and Jianguo Li for his helpful discussions. We also thank Bryce Littlejohn discussions on EOS.
B.~D. and W.-C. Huang are supported by the U.S. Department of Energy (DOE) Grant No. DE-SC0010813. B.~S. Hu is supported by the DOE, Office of Science, under SciDAC-5 (NUCLEI collaboration) and Cyclotron Institute at Texas A\&M University.
We acknowledge support from the Texas A\&M University System National Laboratories Office and Los Alamos National Laboratory. This research used resources from the Oak Ridge Leadership Computing Facility located at Oak Ridge National Laboratory, which is supported by the Office of Science of the DOE under contract No. DE-AC05-00OR22725.
\end{acknowledgments}

\appendix
\section{Multipole expansion}
Nuclear $X$ decay rate (Eq.~\ref{eq:decay}) can be characterized by nuclear matrix element

\begin{align}
    \Gamma_X = \lvert \bra{J_f, X} \hat{H} \ket{J_i} \rvert ^2= \int d^{3}\vec{r} \epsilon_{\alpha}^{\mu} (\vec{q})^* \mathcal{J}_{\mu}\!(\vec{r}) e^{-i\vec{q} \cdot \vec{r}} \label{eq:nme}
\end{align}
where $q$/$\alpha$ are momentum/polarization, $J_i$/$J_f$ are initial/final nuclear state. 
We can expand Eq.~\ref{eq:nme} by Wigner-Eckart Theorem and multipole operators, which are defined by
\begin{eqnarray}
    \hat{\mathcal{M}}_{JM} &=& \int d^3x [j_J(qx)Y_{JM}(\Omega_x)] \hat{\mathcal{J}}_0(x)\\
    \hat{\mathcal{L}}_{JM} &=& \frac{i}{q} \int d^3x \left[ \nabla[j_J(qx)Y_{JM}(\Omega_x)] \right]\cdot \hat{\mathcal{J}}(x) \\
    \hat{\mathcal{T}}_{JM}^{\rm el} &=& \frac{1}{q} \int d^3x [\nabla \times j_J(qx) \textbf{Y}^M_{JJ1}(\Omega_x)]\cdot \hat{\mathcal{J}}(x)\\
    \hat{\mathcal{T}}_{JM}^{\rm mag} &=& \int d^3x [j_J(qx) \textbf{Y}^M_{JJ1}(\Omega_x)]\cdot \hat{\mathcal{J}}(x)
\end{eqnarray}
where $j_J(qx)$ are Bessel functions, $Y_{JM}(\Omega_x)$ are spherical harmonics and $\textbf{Y}^M_{JJ1}(\Omega_x)$ are vector spherical harmonics. Thus Eq.~\ref{eq:nme} reads
\begin{align}
\label{eq:xrate}
\Gamma_{X} &=\frac{2q}{2J_i+1}\Biggl\{ \sum_{J\geq 0}\left| \braket{J_f || \left[ {q\over m} \mathcal{M}_J-{\omega \over m}\mathcal{L}_J \right] || J_i} \right|^2 \nonumber \\
&+ \sum_{J\geq 1} \left[ | \braket{J_f || \mathcal{T}_{J}^{el}||J_i} |^2 + \left|\braket{J_f||\mathcal{T}_{J}^{mag}||J_i}\right|^2 \right] \Biggr\}
\end{align}

\section{Long wavelength limit}
Expanding the multipole operators in terms of transfer momentum $q$,
\begin{align}
\mathcal{M}_{JM} &\simeq {q^J \over (2J+1)!!} \int d^3x \, x^J Y_{JM}\mathcal{J}^0 (\vec{x}) \\
\mathcal{L}_{JM} &\simeq {-i q^{J-1} \over (2J+1)!!}\int d^3x x^J Y_{JM} \nabla \cdot \vec{\mathcal{J}}(\vec{x}) \\
\mathcal{T}_{JM}^{\rm el} &\simeq {-i q^{J-1} \over (2J+1)!!}\sqrt{J+1 \over J} \int d^3x x^J Y_{JM} \nabla \cdot \vec{\mathcal{J}}(\vec{x}) \\
\mathcal{T}_{JM}^{\rm mag} &\simeq {i q^J \over (2J+1)!!} \sqrt{{J+1 \over J}}\int d^3x \times \left[ {1 \over J+1}\vec{x} \times \vec{\mathcal{J}}(\vec{x}) \right] \\
& \,\, \cdot \nabla x^J Y_{JM} \nonumber
\end{align}

In long wavelength limit ($q \rightarrow 0$), the multipole operators are approximated by
\begin{align}
\mathcal{M}_{00} &\simeq \mathcal{M}_{1M} \simeq 0 \\
\mathcal{L}_{00} &\simeq -{iq \over 3 \sqrt{4\pi}}  \sum_s \epsilon_s (\vec{r}_s \cdot\vec{\sigma}_s ) \nonumber \\
&= -{iq \over 6 \sqrt{4\pi}} [(\epsilon_p +\epsilon_n) \hat{d}_0^\sigma + (\epsilon_p- \epsilon_n) \hat{d}_3^\sigma] \\
\mathcal{L}_{1M} &\simeq {i \over \sqrt{12\pi}} \sum_s \epsilon_s \vec{\sigma}_s \cdot \hat{e}_M = {i \over \sqrt{12\pi}} [\epsilon_p \hat{\sigma}_M^p + \epsilon_n \hat{\sigma}_M^n] \\
\mathcal{T}_{1M}^{\rm el} &\simeq {i \over \sqrt{6 \pi}} \sum_s \epsilon_s \vec{\sigma}_s \cdot \hat{e}_M = {i \over \sqrt{6 \pi}} [\epsilon_p \hat{\sigma}_M^p + \epsilon_n \hat{\sigma}_M^n] \\
\mathcal{T}_{1M}^{\rm mag} &\simeq {iq \over 2\sqrt{6\pi}} \sum_s \epsilon_s (\vec{r}_s \times \vec{\sigma}_s) \cdot \hat{e}_M \nonumber \\
&= {iq \over 4\sqrt{6\pi} } [(\epsilon_p + \epsilon_n) \hat{D}_{0M}^\sigma + (\epsilon_p-\epsilon_n) \hat{D}_{3M}^\sigma]
\end{align}
One can reproduce Eq.~\ref{eq:decay_rate} by combining the approximated multipole operators with Eq.~\ref{eq:xrate} and ignoring high order terms.

\section{Experiment simulation}
\begin{figure}[tbh]
    \centering
    \includegraphics[width=\columnwidth]{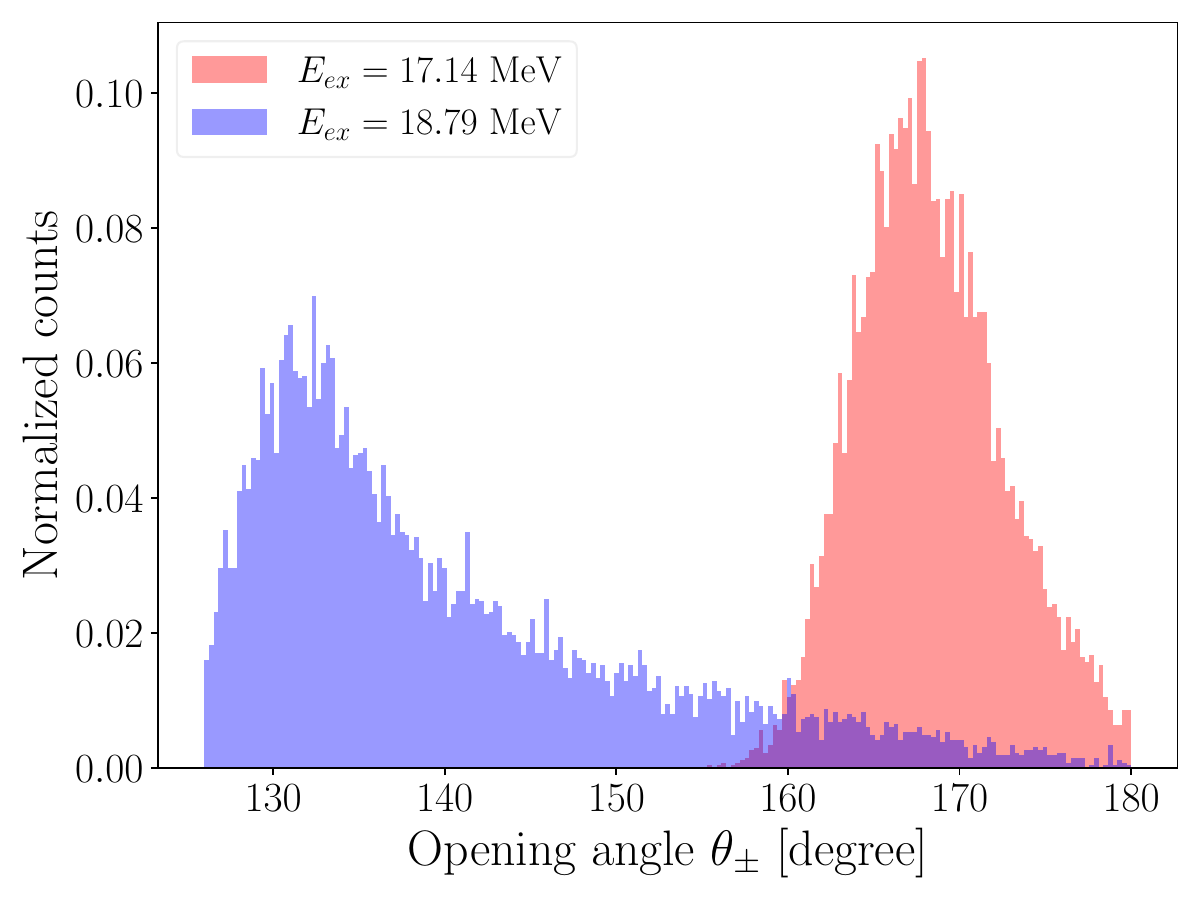}\\
    \includegraphics[width=\columnwidth]{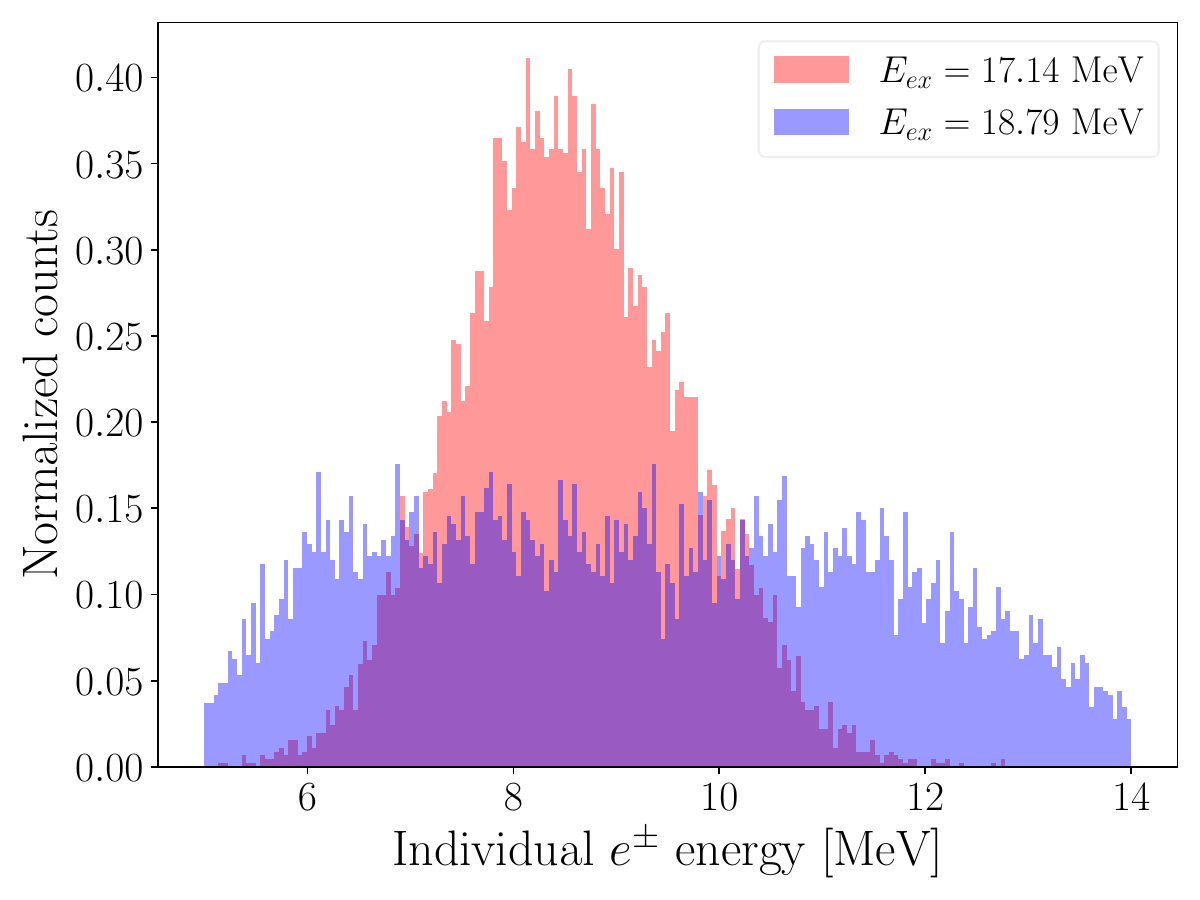}\\
    \caption{Angle (top) and energy (bottom) spectrum of individual $e^\pm$ from a decaying $X$ boson produced by nuclear state $E_{ex}=17.14$ MeV (red) and $E_{ex}=18.79$ MeV (blue). The angle and energy are smeared by gaussian with 3 degree and 10\% energy width, respectively.}
    \label{fig:sim}
\end{figure}

\begin{figure}[tbh]
    \centering
    \includegraphics[width=\columnwidth]{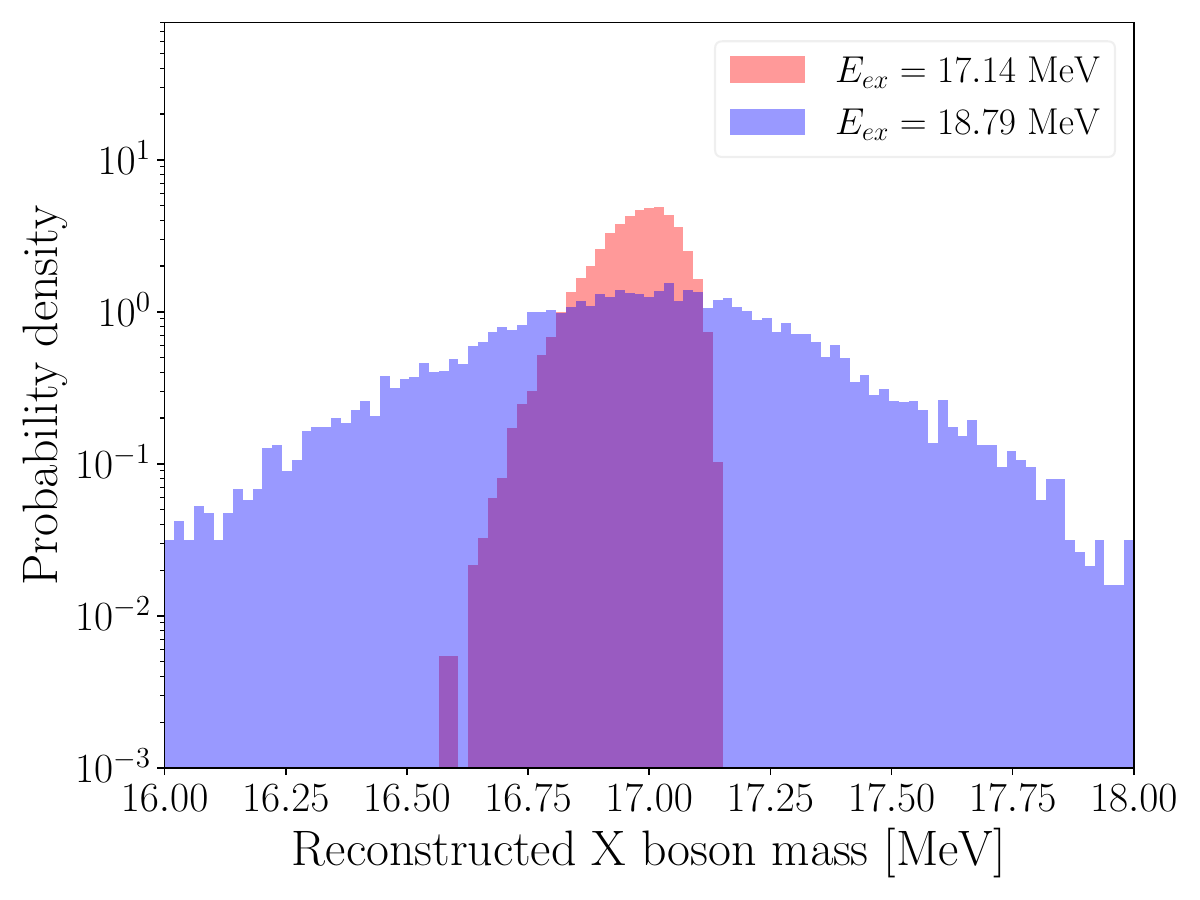}
    \caption{Reconstruct $X$ boson mass from from GEANT simulation Fig.~\ref{fig:sim}. (10\% energy resolution)}
    \label{fig:rec}
\end{figure}

We perform a simulation of $X$ boson (17 MeV mass) decay inside a 130 kg glass sphere 24\,cm in radius at the center of the LAr detector.
$X$ boson is produced by nuclear deexcitation from two different energy states: 17.14 MeV and 18.79 MeV.
These two states are the most probable transitions as described in Table.~\ref{tab:states}.
Fig.~\ref{fig:sim} shows the angle and energy spectrum smeared with detector resolution of 3 degree and 10\% energy, respectively, that results from Cherenkov light reconstruction of the individual $e^+e^-$ leptons. $X$ boson generated from the 17.14 MeV nuclear state has lower kinetic energy 0.14 MeV; thus its angle spectrum is closer to back-to-back decay. $X$ boson with 18.79 MeV has 1.79 MeV kinetic energy, so its angle spectrum is more farther from back-to-back decay. 
The unsmeared (ie. perfect resolution) energy spectrum is flat distribution bounded by the kinematic energy limits (ie. forward/backward decay).
The smearing will make the measured energy outside the limits.
Fig.~\ref{fig:rec} uses energy and angle spectrum in Fig.~\ref{fig:sim} to reconstruct $X$ boson mass $m_X$.
Since $m_X=17$ MeV is assumed in the simulation, the reconstruct mass spectrum peaks at 17 MeV.  Since Cherenkov light is used for reconstruction, backgrounds arising from protons, alphas, deuteriums etc. generated from neutron scattering are small since they only produce scintillation light.  The dominant background will be from mis-id single electrons or photons from random backgrounds, which, above 10 MeV are shown to be small~\cite{CCM:2021jmk}.

\bibliographystyle{bibi}
\bibliography{refs}

\providecommand{\href}[2]{#2}\begingroup\raggedright\begin{thebibliography}{10}

\bibitem{Feng:2016jff}
J.~L. Feng, B.~Fornal, I.~Galon, S.~Gardner, J.~Smolinsky, T.~M.~P. Tait and
  P.~Tanedo, \emph{{Protophobic Fifth-Force Interpretation of the Observed
  Anomaly in $^8$Be Nuclear Transitions}},
  \href{https://doi.org/10.1103/PhysRevLett.117.071803}{\emph{Phys. Rev. Lett.}
  {\bfseries 117} (2016) 071803}
  [\href{https://arxiv.org/abs/1604.07411}{{\ttfamily 1604.07411}}].

\bibitem{Cartlidge2016}
E.~Cartlidge, \emph{Has a hungarian physics lab found a fifth force of
  nature?}, \href{https://doi.org/10.1038/nature.2016.19957}{\emph{Nature}
  (2016) }.

\bibitem{Krasznahorkay:2015iga}
A.~J. Krasznahorkay et~al., \emph{{Observation of Anomalous Internal Pair
  Creation in Be8 : A Possible Indication of a Light, Neutral Boson}},
  \href{https://doi.org/10.1103/PhysRevLett.116.042501}{\emph{Phys. Rev. Lett.}
  {\bfseries 116} (2016) 042501}
  [\href{https://arxiv.org/abs/1504.01527}{{\ttfamily 1504.01527}}].

\bibitem{Krasznahorkay_2018}
A.~J. Krasznahorkay, M.~Csatlos, L.~Csige, Z.~Gacsi, J.~Gulyas, A.~Nagy,
  N.~Sas, J.~Timar, T.~G. Tornyi, I.~Vajda and A.~J. Krasznahorkay, \emph{New
  results on the 8be anomaly},
  \href{https://doi.org/10.1088/1742-6596/1056/1/012028}{\emph{Journal of
  Physics: Conference Series} {\bfseries 1056} (2018) 012028}.

\bibitem{PhysRevC.104.044003}
A.~J. Krasznahorkay, M.~Csatl\'os, L.~Csige, J.~Guly\'as, A.~Krasznahorkay,
  B.~M. Nyak\'o, I.~Rajta, J.~Tim\'ar, I.~Vajda and N.~J. Sas, \emph{New
  anomaly observed in $^{4}\mathrm{He}$ supports the existence of the
  hypothetical x17 particle},
  \href{https://doi.org/10.1103/PhysRevC.104.044003}{\emph{Phys. Rev. C}
  {\bfseries 104} (2021) 044003}.

\bibitem{Krasznahorkay:2019lyl}
A.~J. Krasznahorkay et~al., \emph{{New evidence supporting the existence of the
  hypothetic X17 particle}},
  \href{https://arxiv.org/abs/1910.10459}{{\ttfamily 1910.10459}}.

\bibitem{Krasznahorkay:2022pxs}
A.~J. Krasznahorkay et~al., \emph{{New anomaly observed in C12 supports the
  existence and the vector character of the hypothetical X17 boson}},
  \href{https://doi.org/10.1103/PhysRevC.106.L061601}{\emph{Phys. Rev. C}
  {\bfseries 106} (2022) L061601}
  [\href{https://arxiv.org/abs/2209.10795}{{\ttfamily 2209.10795}}].

\bibitem{Barducci:2022lqd}
D.~Barducci and C.~Toni, \emph{{An updated view on the ATOMKI nuclear
  anomalies}}, \href{https://doi.org/10.1007/JHEP02(2023)154}{\emph{JHEP}
  {\bfseries 02} (2023) 154}
  [\href{https://arxiv.org/abs/2212.06453}{{\ttfamily 2212.06453}}]. [Erratum:
  JHEP 07, 168 (2023)].

\bibitem{Feng:2016ysn}
J.~L. Feng, B.~Fornal, I.~Galon, S.~Gardner, J.~Smolinsky, T.~M.~P. Tait and
  P.~Tanedo, \emph{{Particle physics models for the 17 MeV anomaly in beryllium
  nuclear decays}},
  \href{https://doi.org/10.1103/PhysRevD.95.035017}{\emph{Phys. Rev. D}
  {\bfseries 95} (2017) 035017}
  [\href{https://arxiv.org/abs/1608.03591}{{\ttfamily 1608.03591}}].

\bibitem{Feng:2020mbt}
J.~L. Feng, T.~M.~P. Tait and C.~B. Verhaaren, \emph{{Dynamical Evidence For a
  Fifth Force Explanation of the ATOMKI Nuclear Anomalies}},
  \href{https://doi.org/10.1103/PhysRevD.102.036016}{\emph{Phys. Rev. D}
  {\bfseries 102} (2020) 036016}
  [\href{https://arxiv.org/abs/2006.01151}{{\ttfamily 2006.01151}}].

\bibitem{Kozaczuk:2016nma}
J.~Kozaczuk, D.~E. Morrissey and S.~R. Stroberg, \emph{{Light axial vector
  bosons, nuclear transitions, and the $^8$Be anomaly}},
  \href{https://doi.org/10.1103/PhysRevD.95.115024}{\emph{Phys. Rev. D}
  {\bfseries 95} (2017) 115024}
  [\href{https://arxiv.org/abs/1612.01525}{{\ttfamily 1612.01525}}].

\bibitem{PhysRevD.106.012001}
{\scshape CCM Collaboration} Collaboration, A.~A. Aguilar-Arevalo et~al.,
  \emph{First dark matter search results from coherent captain-mills},
  \href{https://doi.org/10.1103/PhysRevD.106.012001}{\emph{Phys. Rev. D}
  {\bfseries 106} (2022) 012001}.

\bibitem{CCM:2021yzc}
{\scshape CCM} Collaboration, A.~A. Aguilar-Arevalo et~al., \emph{{First
  Leptophobic Dark Matter Search from the Coherent\textendash{}CAPTAIN-Mills
  Liquid Argon Detector}},
  \href{https://doi.org/10.1103/PhysRevLett.129.021801}{\emph{Phys. Rev. Lett.}
  {\bfseries 129} (2022) 021801}
  [\href{https://arxiv.org/abs/2109.14146}{{\ttfamily 2109.14146}}].

\bibitem{CCM:2021jmk}
{\scshape CCM} Collaboration, A.~A. Aguilar-Arevalo et~al., \emph{{Prospects
  for detecting axionlike particles at the Coherent CAPTAIN-Mills experiment}},
  \href{https://doi.org/10.1103/PhysRevD.107.095036}{\emph{Phys. Rev. D}
  {\bfseries 107} (2023) 095036}
  [\href{https://arxiv.org/abs/2112.09979}{{\ttfamily 2112.09979}}].

\bibitem{Aguilar-Arevalo:2023kvr}
A.~A. Aguilar-Arevalo et~al., \emph{{Testing meson portal dark sector solutions
  to the MiniBooNE anomaly at the Coherent CAPTAIN Mills experiment}},
  \href{https://doi.org/10.1103/PhysRevD.109.095017}{\emph{Phys. Rev. D}
  {\bfseries 109} (2024) 095017}
  [\href{https://arxiv.org/abs/2309.02599}{{\ttfamily 2309.02599}}].

\bibitem{Toups:2022yxs}
M.~Toups et~al., \emph{{PIP2-BD: GeV Proton Beam Dump at Fermilab's PIP-II
  Linac}},  in \emph{{Snowmass 2021}}, 3, 2022,
  \href{https://arxiv.org/abs/2203.08079}{{\ttfamily 2203.08079}}.

\bibitem{Anderson:2022lbb}
T.~Anderson et~al., \emph{{Eos: conceptual design for a demonstrator of hybrid
  optical detector technology}},
  \href{https://doi.org/10.1088/1748-0221/18/02/P02009}{\emph{JINST} {\bfseries
  18} (2023) P02009} [\href{https://arxiv.org/abs/2211.11969}{{\ttfamily
  2211.11969}}].

\bibitem{MEGII:2024urz}
{\scshape MEG II} Collaboration, K.~Afanaciev et~al., \emph{{Search for the X17
  particle in $^{7}\mathrm{Li}(\mathrm{p},\mathrm{e}^+ \mathrm{e}^{-})
  ^{8}\mathrm{Be}$ processes with the MEG II detector}},
  \href{https://arxiv.org/abs/2411.07994}{{\ttfamily 2411.07994}}.

\bibitem{Zhang:2020ukq}
X.~Zhang and G.~A. Miller, \emph{{Can a protophobic vector boson explain the
  ATOMKI anomaly?}},
  \href{https://doi.org/10.1016/j.physletb.2021.136061}{\emph{Phys. Lett. B}
  {\bfseries 813} (2021) 136061}
  [\href{https://arxiv.org/abs/2008.11288}{{\ttfamily 2008.11288}}].

\bibitem{Richard:2023int}
R.~Van De~Water, \emph{Probing the dark sector with accelerators: New
  opportunities!},
  \url{https://www.int.washington.edu/sites/default/files/schedule_session_files/VanDeWater_R.pdf},
  April, 2023.
\newblock Talk at Seattle: Interplay of Nuclear, Neutrino and BSM Physics at
  Low-Energies \url{https://www.int.washington.edu/program/schedule/1205}.

\bibitem{r5912}
\emph{Hamamatsu r5912},
\newblock https://hep.hamamatsu.com/eu/en/products/R5912.html.

\bibitem{TILLEY19931}
D.~Tilley, H.~Weller and C.~Cheves, \emph{Energy levels of light nuclei a =
  16–17},
  \href{https://doi.org/https://doi.org/10.1016/0375-9474(93)90073-7}{\emph{Nuclear
  Physics A} {\bfseries 564} (1993) 1}.

\bibitem{Pellico:2022dju}
W.~Pellico, C.~Bhat, J.~Eldred, C.~Johnstone, J.~Johnstone, K.~Seiya, C.-Y.
  Tan, M.~Toups, P.~deNiverville and R.~Van De~Water, \emph{{FNAL PIP-II
  Accumulator Ring}},  \href{https://arxiv.org/abs/2203.07339}{{\ttfamily
  2203.07339}}.

\bibitem{Aguilar-Arevalo:2023dai}
A.~A. Aguilar-Arevalo et~al., \emph{{Physics Opportunities at a Beam Dump
  Facility at PIP-II at Fermilab and Beyond}},
  \href{https://arxiv.org/abs/2311.09915}{{\ttfamily 2311.09915}}.

\bibitem{KELLEY201771}
J.~Kelley, J.~Purcell and C.~Sheu, \emph{Energy levels of light nuclei a=12},
  \href{https://doi.org/https://doi.org/10.1016/j.nuclphysa.2017.07.015}{\emph{Nuclear
  Physics A} {\bfseries 968} (2017) 71}.

\bibitem{Johnson:2018hrx}
C.~W. Johnson, W.~E. Ormand, K.~S. McElvain and H.~Shan, \emph{{BIGSTICK: A
  flexible configuration-interaction shell-model code}},
  \href{https://arxiv.org/abs/1801.08432}{{\ttfamily 1801.08432}}.

\bibitem{Johnson:2013bna}
C.~W. Johnson, W.~E. Ormand and P.~G. Krastev, \emph{{Factorization in
  large-scale many-body calculations}},
  \href{https://doi.org/10.1016/j.cpc.2013.07.022}{\emph{Comput. Phys. Commun.}
  {\bfseries 184} (2013) 2761}
  [\href{https://arxiv.org/abs/1303.0905}{{\ttfamily 1303.0905}}].

\bibitem{PhysRevC.46.923}
E.~K. Warburton and B.~A. Brown, \emph{Effective interactions for the 0p1s0d
  nuclear shell-model space},
  \href{https://doi.org/10.1103/PhysRevC.46.923}{\emph{Phys. Rev. C} {\bfseries
  46} (1992) 923}.

\bibitem{Yuan:2012zz}
C.~Yuan, T.~Suzuki, T.~Otsuka, F.~Xu and N.~Tsunoda, \emph{{Shell-model study
  of boron, carbon, nitrogen, and oxygen isotopes with a monopole-based
  universal interaction}},
  \href{https://doi.org/10.1103/PhysRevC.85.064324}{\emph{Phys. Rev. C}
  {\bfseries 85} (2012) 064324}
  [\href{https://arxiv.org/abs/1209.5587}{{\ttfamily 1209.5587}}].

\bibitem{michel2021gamow}
N.~Michel and M.~P{\l}oszajczak, \emph{Gamow Shell Model: The Unified Theory of
  Nuclear Structure and Reactions}. Springer International Publishing, 2021.

\bibitem{PhysRevC.106.L011301}
N.~Michel, J.~G. Li, L.~H. Ru and W.~Zuo, \emph{Calculation of the
  thomas-ehrman shift in $^{16}\mathrm{F}$ and $^{15}\mathrm{O}(p,p)$ cross
  sections within the gamow shell model},
  \href{https://doi.org/10.1103/PhysRevC.106.L011301}{\emph{Phys. Rev. C}
  {\bfseries 106} (2022) L011301}.

\bibitem{PTP.62.981}
H.~Furutani, H.~Horiuchi and R.~Tamagaki, \emph{{Cluster-Model Study of the T=1
  States in A=4 System: $^3$He+p Scattering}},
  \href{https://doi.org/10.1143/PTP.62.981}{\emph{Prog. Theor. Phys.}
  {\bfseries 62} (1979) 981}.

\bibitem{PTPS.68.193}
H.~Furutani, H.~Kanada, T.~Kaneko, S.~Nagata, H.~Nishioka, S.~Okabe, S.~Saito,
  T.~Sakuda and M.~Seya, \emph{{Chapter III. Study of Non-Alpha-Nuclei Based on
  the Viewpoint of Cluster Correlations}},
  \href{https://doi.org/10.1143/PTPS.68.193}{\emph{Prog. Theor. Phys. Suppl.}
  {\bfseries 68} (1980) 193}.

\bibitem{PhysRevC.96.054316}
Y.~Jaganathen, R.~M.~I. Betan, N.~Michel, W.~Nazarewicz and M.~P\l{}oszajczak,
  \emph{Quantified gamow shell model interaction for $psd$-shell nuclei},
  \href{https://doi.org/10.1103/PhysRevC.96.054316}{\emph{Phys. Rev. C}
  {\bfseries 96} (2017) 054316}.

\bibitem{Jianguo}
J.~G. Li, \emph{Private communication}, .

\bibitem{Darcy:2024pheno}
D.~Newmark, \emph{Cherenkov light identification at coherent captain-mills
  experiment},
  \url{https://indico.global/event/805/contributions/23562/attachments/11187/16552/MayDPF2024%20(1).pdf},
  May, 2024.
\newblock DPF-PHENO (to be published).

\end{thebibliography}\endgroup

\end{document}